\documentclass[twoside]{article}
\usepackage{fleqn,espcrc2,epsfig}

\newcommand{\be}{\begin{equation}}                                              
\newcommand{\ee}{\end{equation}}

\newcommand{\AmS}{{\protect\the\textfont2
  A\kern-.1667em\lower.5ex\hbox{M}\kern-.125emS}}

\title{Simulation of QCD and other similar theories}

\author{I. Montvay \address{Deutsches Elektronen Synchrotron, DESY,
                            Notkestr. 85, D-22603 Hamburg, Germany}}

\begin{document}

\begin{abstract}
 In QCD one can change the representation of the gauge group for quarks
 and/or the gauge group itself.
 Examples of such generalizations are: (a) supersymmetric Yang-Mills
 theory with gauge group SU(2) or SU(3); (b) QCD with SU(2) colour and
 quarks in the adjoint representation; (c) QCD with three equal-mass
 quarks.
 Experience and perspectives in simulations of these theories by
 applying the two-step multi-bosonic algorithm is described.
 As an example of an interesting physical problem in these theories the
 expected spontaneous CP-violation in QCD with three equal-mass quarks
 is discussed.
\end{abstract}

\maketitle

\section{INTRODUCTION}\label{sec1}
 Quantum chromodynamics is most probably the correct theory of strong
 interactions.
 It belongs to a class of relativistic quantum field theories in four
 dimensions which are believed to describe a wide range of physical
 phenomena consistently from zero to infinite energies.

 Examples of generalizations of QCD are given in the abstracts.
 Numerical simulations of these theories can give important
 non-perturbative insight in low energy dynamical features as
 confinement and chiral symmetry breaking.
 Restricting numerical studies to the quenched approximation is
 obviously unsatisfactory.
 In QCD, besides dynamical simulations with two equal mass quarks, the
 investigation of $\bf 2 \oplus 1$ light dynamical quark flavours is
 very important because it is a good approximation of the physically
 relevant $u$-,$d$- and $s$-quark system.

 A fermion simulation algorithm which is able to deal with QCD-like
 theories for small fermion masses in sufficiently large volumes is
 the two-step multi-bosonic (TSMB) algorithm \cite{TSMB,POLYNOM}.
 In a recent large-scale simulation \cite{CHIRAL,SPECTRUM,DESYMUNSTER}
 this algorithm has been used for the numerical study of supersymmetric
 Yang-Mills theories which contain ``gauginos'' i.e.\ Majorana fermions
 in the adjoint representation of the gauge group.
 The algorithm could cope with a potentially difficult situation where
 massless gauginos are playing an important dynamical r\^ole in
 reasonably large physical volumes.
 It is an interesting question whether this experience can be extended,
 for instance, to QCD with $\bf 2 \oplus 1$ light dynamical quarks.

\section{FERMION ALGORITHM}\label{sec2}
 The TSMB algorithm is based on the multi-bosonic representation of
 the fermion determinant \cite{LUSCHER}:
\be\label{eq01}
\left|\det(Q)\right|^{N_f} \simeq \frac{1}{\det P_n(Q^\dagger Q)} \ .
\ee
 Here $Q$ is the fermion matrix and $N_f$ degenerate flavours are
 considered. 
 The polynomial $P_n(x)$ is an approximation of the function
 $x^{-\alpha},\;(\alpha\equiv N_f/2)$ in an interval
 $[\epsilon,\lambda]$ covering the spectrum of $Q^\dagger Q$:
\be\label{eq02}
\lim_{n \to \infty} P_n(x) = x^{-N_f/2} \hspace{2em}{\rm for} 
\hspace{1em} x \in [\epsilon,\lambda] \ .
\ee
 The inverse determinant of $P_n$ in (\ref{eq01}) can be represented by
 $n$ bosonic ``pseudofermion'' fields.
 Since for a good approximation a large number ($n={\cal O}(1000)$) of
 pseudofermion fields would be necessary the direct application of
 eq.~(\ref{eq01}) to numerical simulations is not practical.
 In TSMB a two-step approximation by a product of two polynomials is
 introduced
\be\label{eq03}
\lim_{n_2 \to \infty} P^{(1)}_{n_1}(x)P^{(2)}_{n_2}(x) = 
x^{-N_f/2} \ .
\ee
 The multi-bosonic representation is applied for the first polynomial
 $P^{(1)}_{n_1}$ and the determinant of the second one is taken into
 account in a stochastic ``noisy correction'' step.
 For very small fermion masses it is advantageous to perform a final
 ``reweighting correction'' in the process of evaluating the expectation
 values.
 (For a comprehensive summary of algorithmic details and for references
 see \cite{SPECTRUM}.)

 The DESY-M\"unster collaboration collected extensive experience on the
 behaviour of the TSMB algorithm in an application to supersymmetric
 Yang-Mills (SYM) theory \cite{CHIRAL,SPECTRUM,DESYMUNSTER}.
 The total amount of computer time used in last year corresponds to
 $\sim 0.01\, {\rm Tflopyear}$ (sustained performance with 64 bit
 precision).
 The largest lattice was $12^3 \cdot 24$.
 In QCD units obtained from the string tension the spatial extension
 of this lattice is $\sim 1\, {\rm fm}$.
 The statistics was relatively high: several thousands independent gauge
 configurations per simulation point.
 Since a Majorana fermion is equivalent to half of a Dirac fermion, the
 gaugino corresponds to 
\be\label{eq04}
\alpha=N_f/2=1/4 \ .
\ee

 An important factor in the extrapolation to QCD is the dependence on
 the number of flavours $N_f$.
 It turned out that both execution time of a sweep and autocorrelation
 of important physical quantities are roughly proportional to the number
 of pseudofermion fields $n_1$.
 This has to be chosen such that the acceptance rate in the noisy
 correction step stays high enough.
 (In our simulations we always had 85-90\% acceptance.)
 This can be achieved if the deviation norm \cite{POLYNOM} satisfies 
\be\label{eq05}
\delta_{n_1}^2 \leq 5 \cdot 10^{-5} \ .
\ee
 On the basis of this it is possible to estimate the dependence on the
 number of flavours because, in the limit of a small lower limit of the
 approximation interval $\epsilon \ll \lambda$, the relation
 \cite{POLYNOM} 
\be\label{eq06}
\delta_n(\alpha;\epsilon=0,\lambda=1) =
\frac{\alpha}{n+1+\alpha}\; \simeq\; \frac{N_f}{2n}
\ee
 gives a good estimate.
 This shows that the number of necessary pseudofermion fields $n_1$ is
 approximately proportional to $N_f$ and the cost for obtaining a new
 independent gauge configuration to $N_f^2$.

 Let us note in this context that in a future QCD simulation one would
 like to have, for instance, a lattice spacing
 $a \simeq (4\, {\rm GeV})^{-1}$ and a light quark mass 
 $m_q \simeq 4\, {\rm MeV}$ implying $am_q \simeq 10^{-3}$.
 This would correspond to $\epsilon \simeq 10^{-6}$, a value which has
 also been exploited in SYM simulations \cite{CHIRAL}.

 Most of the experience on the TSMB algorithm is concentrated up to now
 on Wilson fermions.
 The implementation of ``moderately improved'' actions, as e.g.\ the
 clover action of Sheikholeslami-Wohlert, seems possible with reasonable
 effort.
 Recently the algorithm has also been applied for staggered fermions
 \cite{CHEMICAL}.

 An important question for numerical simulations of odd numbers of
 quark flavours with Wilson fermions is the effect of the sign of the
 fermion determinant.
 In eq.~(\ref{eq01}) the absolute value of the (real) determinant
 appears therefore the sign has to be taken into account as a
 reweighting factor during the evaluation of expectation values.
 This is a potential problem in at least two respects: the computation
 of the determinant is only possible on very small lattices and even
 if the sign of the determinant is known the reweighting with it can
 in principle cause a serious loss of accuracy due to cancellations.
 In SYM the problem is similar but the r\^ole of the determinant is
 played by the Pfaffian resulting from the Grassmann integration for
 Majorana fermions.

 Concerning the first problem a satisfactory solution is to monitor the
 sign of the determinant (or Pfaffian) by the {\em spectral flow}
 method \cite{SPECTRUM}.
 The determinant (or Pfaffian) has a positive sign at small enough
 hopping parameters (say, $K \leq 0.125$) and one can count the number
 of sign changes up to the required (dynamical) hopping parameter by
 monitoring the flow of the eigenvalues of the hermitean fermion matrix
 $\tilde{Q}\equiv\gamma_5 Q$ across zero.
 The problem of strong cancellations among contributions with opposite
 sign remains a potential danger.

 In SYM with SU(2) gauge group it turned out, however, that the
 configurations with positive Pfaffian strongly dominate
 \cite{SPECTRUM}.
 In fact for smaller values of the hopping parameter $K$ no
 configurations with negative Pfaffian were found.
 Negative signs at the percent level occurred only for $K$-values 
 beyond $K=K_0$ corresponding to zero gaugino mass.

 The sign of the determinant in QCD with $\bf 2 \oplus 1$ light flavours
 was up to now not investigated.
 However, the $s$-quark is presumably not light enough for this problem
 to become serious.

\section{THREE-FLAVOUR QCD}\label{sec3}
 QCD with three equal mass quarks is not realized in Nature but from the
 theoretical point of view is rather interesting.
 An important feature is the possibility of {\em spontaneous 
 CP-violation} at negative quark mass predicted long ago by Dashen
 \cite{DASHEN} and Witten \cite{WITTEN}.
 This is a very interesting theoretical issue especially in view of
 recent and near-future CP-violation experiments in B-factories.
 In the lattice literature the question of spontaneous CP- and
 flavour-symmetry violation has been extensively studied but mainly
 in case of $N_f=2$ flavours \cite{AOKI,CREUTZ} where it can also be
 just a lattice artifact \cite{SHARPESINGLE}.
 A similar phenomenon also occurs in other models, for instance, in SYM
 theory with SU(3) gauge group (see \cite{DESYMUNSTER}).

\begin{figure}[th]
\vspace*{-0.0cm}
\begin{center}
\epsfig{file=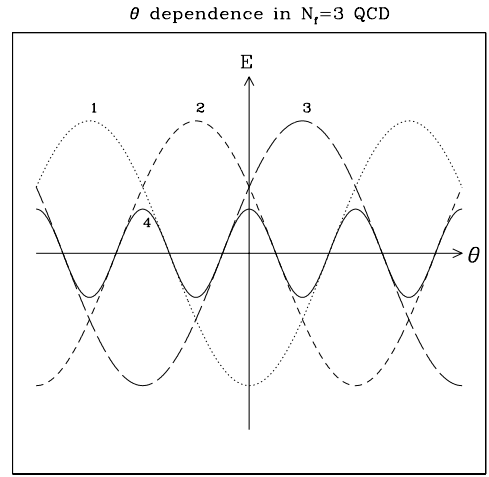,
        width=7.5cm,height=7.0cm,
        angle=0}
\vspace*{-1.8cm}
\caption{\label{fig01}
 The energies of different branches of ground states as a function
 of the $\theta$-parameter.}
\end{center}
\vspace*{-0.7cm}
\end{figure}

 Spontaneous CP-violation in QCD with three flavours is predicted by the
 analysis of the low energy effective chiral Lagrangian
\begin{eqnarray}\label{eq07}
{\cal L} &=& \frac{f_\pi^2}{2} \left[
{\rm Tr\,}\left( \partial_\mu U\, \partial_\mu U^{-1} \right) \right.
\nonumber \\
&+& \left. 2\,{\rm Re\,Tr\,}
\left(m_q e^{-i\theta/3} U \right)\right] \ .
\end{eqnarray}
 Here $f_\pi$ is the usual parameter of chiral dynamics, $U \in SU(3)$
 and $\theta$ is the parameter governing CP-violation.
 At $\theta \ne 0,\pi$ there is explicit CP-violation whereas at
 $\theta=3\pi\equiv\pi$ corresponding to negative quark mass spontaneous
 CP-violation is expected.
 The $\theta$-dependence of the energies of different ground states are
 illustrated by fig.~\ref{fig01}.

 This picture is the solution of the puzzle emphasized in \cite{WITTEN}:
 the vacuum energy in the $N_c\to\infty$ limit is
\be\label{eq08}
E(\theta) = N_c^2 h\left(\frac{\theta}{N_c}\right) = E(\theta+2\pi) \ .
\ee
 The solution can only be a multi-branched function
\be\label{eq09}
E(\theta) = N_c^2\; \min_k\;  h\left(\frac{\theta+2k\pi}{N_c}\right) \ .
\ee
 Numerical simulations could prove this picture following from low
 energy effective chiral Lagrangians.


 {\bf Acknowledgement:} It is a pleasure to thank the members of the
 DESY-M\"unster collaboration for helpful discussions.




\begin{thebibliography}{99}
%
\bibitem{TSMB}
I. Montvay,
Nucl. Phys. B466 (1996) 259.
%
\bibitem{POLYNOM}
I. Montvay,
Comput. Phys. Commun. 109 (1998) 144.
%
\bibitem{CHIRAL}
R. Kirchner, S. Luckmann, I. Montvay, K. Spanderen, J. Westphalen,
Phys. Lett. B446 (1999) 209.
%
\bibitem{SPECTRUM}
I. Campos, A. Feo, R. Kirchner, S. Luckmann, I. Montvay, G. M\"unster, 
K. Spanderen, J. Westphalen,
hep-lat/9903014, to appear in Eur. Phys. J. C.
%
\bibitem{DESYMUNSTER}
A. Feo, R. Kirchner, S. Luckmann, I. Montvay, G. M\"unster, A. Vladikas,
this Proceedings.
%
\bibitem{LUSCHER}
M. L\"uscher, 
Nucl. Phys. B418 (1994) 637.
%
\bibitem{CHEMICAL}
S. Hands, I. Montvay, S. Morrison, M. Oevers, J. Skullerud,
in preparation.
%
\bibitem{DASHEN}
R. Dashen,
Phys. Rev. D3 (1971) 1879.
%
\bibitem{WITTEN}
E. Witten,
Annals of Phys. 128 (1980) 363.
%
\bibitem{AOKI}
S. Aoki,
Phys. Rev. D30 (1984) 2653; \\
Phys. Rev. Lett. 57 (1986) 3136.
%
\bibitem{CREUTZ}
M. Creutz,
Phys. Rev. D52 (1995) 2951.
%
\bibitem{SHARPESINGLE}
S. Sharpe, R. Singleton,
Phys. Rev. D58 (1998) 074501.
%
\end{thebibliography}
\end{document}